\documentstyle[12pt]{article}
\newcommand{\Z}{{\sf Z \!\!\! Z}}

\input{psfig}
\makeatletter
\@addtoreset{equation}{section}
\makeatother
 
\title{Complete Wetting of Gluons and Gluinos
\footnote{This work is supported in part by funds provided by the U.S.
Department of Energy (D.O.E.) under cooperative research agreement
DE-FC02-94ER40818.}}
\author{A. Campos, K. Holland and U.-J. Wiese \\ \\
Center for Theoretical Physics, \\
Laboratory for Nuclear Science, and Department of Physics \\
Massachusetts Institute of Technology (MIT) \\
Cambridge, Massachusetts 02139, U.S.A. \\ \\
MIT Preprint, CTP 2768 \\ \\}
 
\begin{document}
\maketitle
\begin{abstract} \normalsize
 
Complete wetting is a universal phenomenon associated with interfaces 
separating coexisting phases. For example, in the pure gluon theory, at $T_c$ 
an interface separating two distinct high-temperature deconfined phases splits 
into two confined-deconfined interfaces with a complete wetting layer of 
confined phase between them. In supersymmetric Yang-Mills theory, distinct 
confined phases may coexist with a Coulomb phase at zero temperature. In that 
case, the Coulomb phase may completely wet a confined-confined interface. 
Finally, at the high-temperature phase transition of gluons and gluinos, 
confined-confined interfaces are completely wet by the deconfined phase, and 
similarly, de\-con\-fined-de\-con\-fined interfaces are completely wet by the 
confined phase. For these various cases, we determine the interface profiles 
and the corresponding complete wetting critical exponents. The exponents depend
on the range of the interface interactions and agree with those of 
corresponding condensed matter systems.

\end{abstract}
 
\maketitle
 
\newpage

\section{Introduction}

Supersymmetric gauge theories have attracted a lot of attention because, for
example, some aspects of their dynamics are accessible to analytic methods. In 
particular, techniques from string theory can be applied to these systems. The 
$SU(N)$ Yang-Mills theory with ${\cal N}=1$ supersymmetry describes the 
interactions between gluons and gluinos. Here we investigate this model both at
zero and at non-zero temperatures. Of course, at non-zero temperatures
supersymmetry is explicitly broken due to different boundary conditions for 
gluons and gluinos in the Euclidean time direction. This is no problem because 
we will use methods of effective field theory that do not rely on 
supersymmetry. At low temperatures, the theory is in a confined phase with a 
spontaneously broken $\Z(N)_\chi$ chiral symmetry, while at high temperatures 
one expects chiral symmetry to be restored and gluons and gluinos to be 
deconfined. In the deconfined phase, the $\Z(N)_c$ center symmetry of the 
$SU(N)$ gauge group is spontaneously broken. In addition, at zero temperature 
gluons and gluinos may exist in a non-Abelian Coulomb phase without confinement
or chiral symmetry breaking. The existence of the Coulomb phase is a 
controversial issue. For the rest of this paper, we assume that it exists. The 
rich phase structure gives rise to interesting effects related to the 
interfaces separating various bulk phases. Due to spontaneous $\Z(N)_\chi$ 
breaking, there are $N$ distinct confined phases with different values of the 
gluino condensate $\chi^{(n)} = \chi_0 \exp(2 \pi i n/N)$, $n \in 
\{1,2,...,N\}$. These phases are separated by confined-confined interfaces. 
Similarly, at high temperatures, $\Z(N)_c$ breaking gives rise to $N$ 
deconfined phases with different values of the Polyakov loop $\Phi^{(n)} = 
\Phi_0 \exp(2 \pi i n/N)$, $n \in \{1,2,...,N\}$, which are separated by 
deconfined-deconfined interfaces. Assuming a first order phase transition, at 
$T_c$ there are, in addition, confined-deconfined interfaces. Finally, at zero 
temperature, the confined phases can coexist with the Coulomb phase, which has 
a vanishing gluino condensate $\chi^{(0)} = 0$. Hence, there are also 
confined-Coulomb interfaces. 

The properties of confined-Coulomb and confined-confined interfaces at zero 
temperature have been studied in detail by Shifman and collaborators 
\cite{Kov97,Kog98}. In particular, the tension of an interface separating two 
bulk phases with gluino condensates $\chi^{(m)}$ and $\chi^{(n)}$ is 
constrained by the inequality
\begin{equation}
\alpha_{mn} \geq \frac{N}{8 \pi^2} |\chi^{(m)} - \chi^{(n)}| =
\frac{N \chi_0}{8 \pi^2} |\exp(2 \pi i m/N) - \exp(2 \pi i n/N)|.
\end{equation}
If the interface represents a BPS saturated state, this inequality is satisfied
as an equality. Indeed, confined-Coulomb interfaces are BPS saturated
\cite{Kog98}, such that
\begin{equation}
\alpha_{0n} = \frac{N \chi_0}{8 \pi^2}.
\end{equation}
The interface tensions determine the shape of a droplet of Coulomb phase that
wets a confined-confined domain wall. As shown in figure 1a, such a droplet
forms a lens with opening angle $\theta$, where
\begin{equation}
\alpha_{mn} = 2 \alpha_{0n} \cos{\frac{\theta}{2}}.
\end{equation}
\begin{figure}[htb]
\psfig{figure=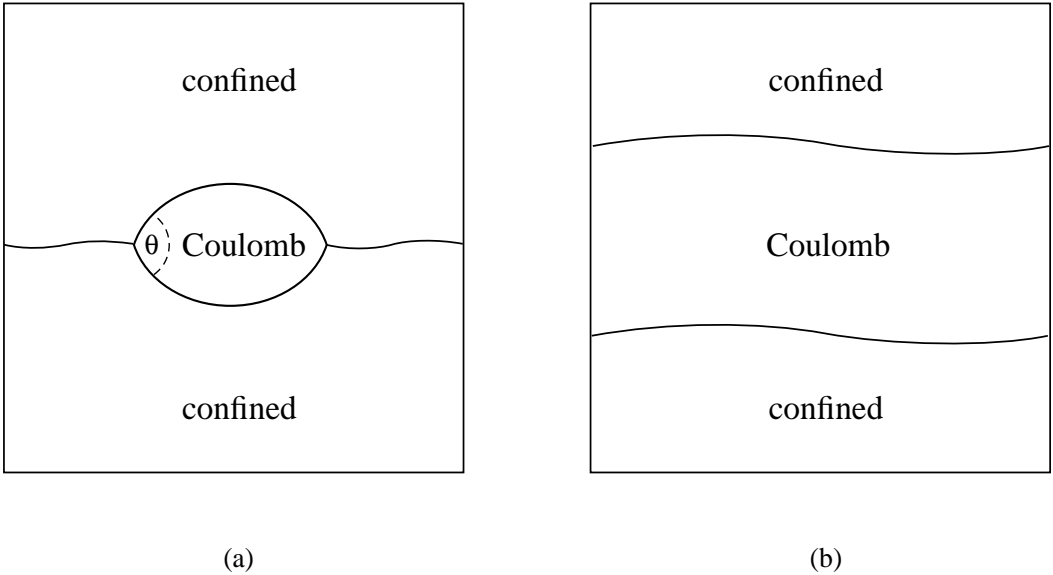,height=3in,width=5.5in}
\caption{\it Incomplete versus complete wetting. (a) For $\alpha_{mn} < 
2 \alpha_{0n}$ one has incomplete wetting with $\theta \neq 0$. Then the 
Coulomb phase forms a lens shaped droplet at the confined-confined interface.
(b) Complete wetting corresponds to $\alpha_{mn} = 2 \alpha_{0n}$. Then
$\theta = 0$, and the Coulomb phase forms a film that splits the
confined-confined domain wall into two confined-Coulomb interfaces.} 
\end{figure}
This equation for $\theta$ follows from the forces at the corner of the lens 
being in equilibrium (forming a stable configuration known as the
Neumann triangle in condensed matter literature \cite{Buf60}). 
In condensed matter physics the inequality 
\begin{equation}
\alpha_{mn} \leq 2 \alpha_{0n},
\end{equation}
was derived by Widom and is also known as Antonoff's rule
\cite{Wid75}. It follows from thermodynamic 
stability because a hypothetical confined-confined domain wall with 
$\alpha_{mn} > 2 \alpha_{0n}$ would simply split into two confined-Coulomb 
interfaces. Combining Antonoff's rule with the BPS inequality, one finds
\begin{equation}
\frac{N \chi_0}{8 \pi^2} |\exp(2 \pi i m/N) - \exp(2 \pi i n/N)| \leq 
\alpha_{mn} \leq \frac{N \chi_0}{4 \pi^2}.
\end{equation}
Assuming BPS saturation for confined-confined interfaces, the values of the 
interface tensions from above imply
\begin{equation}
\theta = \pi - \frac{2 \pi}{N}(m - n), 
\ \mbox{for} \ 1 \leq m-n \leq \frac{N}{2}.
\end{equation}
Then, for odd $N$, $\theta \neq 0$, such that the Coulomb phase forms droplets 
at a confined-confined interface. In condensed matter physics this phenomenon 
is known as incomplete wetting. For even $N$ and for $m-n = N/2$, on the other 
hand, $\alpha_{mn} = 2 \alpha_{0n}$ and $\theta = 0$. In that case, the lens 
shaped droplet degenerates to an infinite film, as shown in figure 1b. When 
such a film is formed, this is called complete wetting. For example, the human 
eye is completely wet by a film of tears that splits the eye-air solid-gas 
interface into a pair of solid-liquid and liquid-gas interfaces. Complete 
wetting may occur when several bulk phases coexist with each other. For 
condensed matter systems, achieving phase coexistence usually requires 
fine-tuning of, for example, the temperature, to a first order phase 
transition. In supersymmetric theories, on the other hand, drastically 
different phases, like confined and Coulomb, may coexist at zero temperature, 
since their bulk free energies are identical due to supersymmetry. This implies
that complete wetting appears naturally, i.e. without fine-tuning, in 
supersymmetric theories. 

Complete wetting is a universal phenomenon of interfaces characterized by
several critical exponents. For example, the width $r$ of the complete
wetting layer diverges as
\begin{equation}
r \propto (T-T_c)^{-\psi},
\end{equation}
where $T-T_c$ measures the deviation from the point of phase coexistence and
$\psi$ is a critical exponent. The value of $\psi$ depends on the range of the
interactions between the two interfaces that enclose the complete wetting
layer. For condensed matter systems, the interaction energy per unit area is
often due to repulsive van der Waals forces and is given by $c/r^p$. In 
addition, away from $T_c$ the phase that forms the wetting layer has a slightly
larger bulk free energy than the other phases. Close to $T_c$, the additional 
bulk free energy per unit area for a wetting layer of width $r$ is 
given by $\mu (T-T_c) r$. Hence, the total free energy per unit area of the two
interface system relative to the free energy of two infinitely separated
interfaces at $T_c$ is given by
\begin{equation}
\alpha_{mn}(T) - 2 \alpha_{0n}(T_c) = \frac{c}{r^p} + \mu (T-T_c) r.
\end{equation}
Minimizing the free energy with respect to $r$, one finds the equilibrium width
\begin{equation}
r \propto (T-T_c)^{-1/(p+1)},
\end{equation}
such that $\psi = 1/(p+1)$. Inserting the equilibrium value of $r$ one finds
\begin{equation}
\alpha_{mn}(T) - 2 \alpha_{0n}(T_c) \propto (T-T_c)^{p/(p+1)} \propto 
(T-T_c)^{1-\psi}.
\end{equation}
Some condensed matter systems with interfaces interact via retarded van der 
Waals forces with a potential $c/r^3$, and thus have $\psi = 1/4$. As we will 
see, the confined-confined interfaces of the ${\cal N} = 1$ supersymmetric 
$SU(N)$ gauge theory, which are completely wet by the non-Abelian Coulomb 
phase, have the same critical exponent.

Complete wetting does not only occur in supersymmetric gauge theories, although
only there it occurs naturally without fine-tuning. In fact, Frei and 
Patk\'{o}s conjectured that complete wetting occurs in the non-supersymmetric 
$SU(3)$ pure gauge theory at the high-temperature deconfinement phase 
transition \cite{Fre89}. In the deconfined phase the $\Z(3)_c$ center symmetry 
is spontaneously broken. Hence, there are three distinct high-temperature 
phases separated by deconfined-deconfined interfaces. When a 
deconfined-deconfined interface is cooled down to the phase transition, the 
low-temperature confined phase forms a complete wetting layer. In that case, 
the interactions between the interfaces are short-ranged, such that
\begin{equation}
\alpha_{mn}(T) - 2 \alpha_{0n}(T_c) = a \exp(- b r) + \mu (T-T_c) r.
\end{equation}
Minimizing the free energy with respect to $r$ now gives
\begin{equation}
r \propto \log (T-T_c),
\end{equation}
such that $\psi = 0$. Substituting the value of $r$, one obtains
\begin{equation}
\alpha_{mn}(T) - 2 \alpha_{0n}(T_c) \propto (T-T_c) \log (T-T_c).
\end{equation}
This is the critical behavior that was observed in ref.\cite{Tra92}. As we will
see, the same exponents follow for the high-temperature deconfinement 
transition in the supersymmetric case.

As argued before, for ${\cal N} = 1$ supersymmetric $SU(3)$ gauge theory (and
generally for odd $N$) complete wetting may occur at zero temperature only if
confined-confined domain walls are not BPS saturated. Recently, we have shown 
that complete wetting occurs at the high-temperature deconfinement phase 
transition \cite{Cam98}. In this case, a confined-confined interface is 
completely wet by one of the three deconfined phases when it is heated up to 
the phase transition. The occurrence of deconfined phase at the center of a 
confined-confined interface has interesting dynamical consequences. It implies 
that a static quark has a finite free energy close to such a domain wall 
and the string emanating from it can end at the wall. This effect was first
derived by Witten in the framework of M-theory \cite{Wit97}. Complete wetting 
provides a field theoretic explanation of the same effect without reference to 
string theory. Here we concentrate on the derivation of the corresponding 
complete wetting critical exponents. As in the non-supersymmetric case, we find
$\psi = 0$ due to short-range interactions between interfaces.

\section{Complete Wetting at Zero Temperature}

In ${\cal N} = 1$ supersymmetric $SU(N)$ gauge theory with even $N$ a 
confined-confined interface separating phases with gluino condensates 
$\chi^{(m)}$ and $\chi^{(n)}$ is completely wet by the non-Abelian Coulomb 
phase if $m - n = N/2$. In this case, Antonoff's rule implies BPS saturation of
confined-confined domain walls. The simplest case we can consider is $SU(2)$, 
where $\chi^{(1)} = \chi_0$ and $\chi^{(2)} = - \chi_0$. The universal aspects 
of the interface dynamics can be derived from an effective action
\begin{equation}
S[\chi] = \int d^4 x \ [\frac{1}{2} \partial_\mu \chi \partial_\mu \chi +
V(\chi)],
\end{equation}
for the gluino condensate $\chi$. Under the $\Z(2)_\chi$ chiral symmetry $\chi$
changes sign and $V(-\chi) = V(\chi)$. When the two confined phases coexist 
with the Coulomb phase, $V(\chi)$ has three degenerate minima, two at $\chi = 
\pm \chi_0$ and one at $\chi = 0$. The confined phase is massive, and thus 
$V(\chi)$ is quadratic around $\chi = \pm \chi_0$. The Coulomb phase, on the 
other hand, is massless and $V(\chi)$ turns out to be quartic around 
$\chi = 0$ \cite{Kog98}. A simple potential with these properties can be 
written as
\begin{equation}
V(\chi) = \chi^4 (a + b \chi^2 + c \chi^4).
\end{equation}
For our purposes it is not essential to use the actual potential of the
supersymmetric theory. We are interested only in the universal complete wetting
aspects of the interface dynamics. These are the same for the simple potential
from above. For $a>0$ there is a quartic minimum at $\chi = 0$ corresponding to
the Coulomb phase. For $0<a<9b^2/32c$ and $b<0$ there are two other minima at
\begin{equation}
\chi_0^2 = \frac{- 3b + \sqrt{9b^2 - 32ac}}{8c},
\end{equation}
corresponding to the two confined phases. Phase coexistence corresponds to 
$b^2 = 4ac$, because then all three minima are degenerate.

We now look for solutions of the classical equation of motion representing
static planar domain walls, i.e. $\chi(x,y,z,t) = \chi(z)$, where $z$ is the 
coordinate perpendicular to the wall. The equation of motion then takes the 
form
\begin{equation}
\frac{d^2 \chi}{dz^2} = \frac{\partial V}{\partial \chi} =
2 \chi^3 (2a + 3b \chi^2 + 4c \chi^4).
\end{equation}
The interface action per unit area and unit time gives the interface tension 
$\alpha_{mn}$. Figure 2 shows a numerical solution of the equation of motion 
for a domain wall separating the two confined phases with boundary conditions 
$\chi(\pm \infty) = \pm \chi_0$. 
\begin{figure}
\psfig{figure=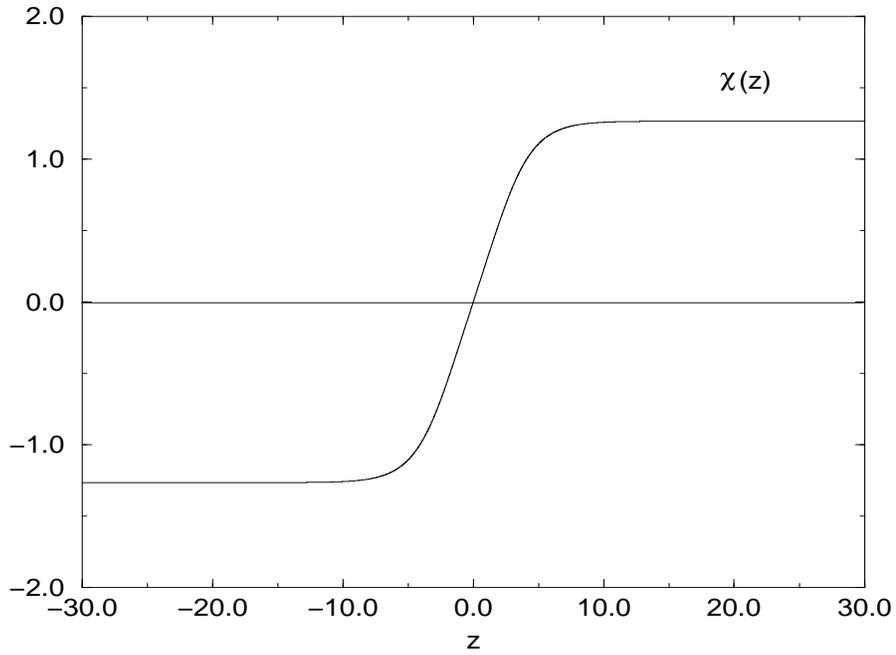,height=4in,width=5.5in}
\psfig{figure=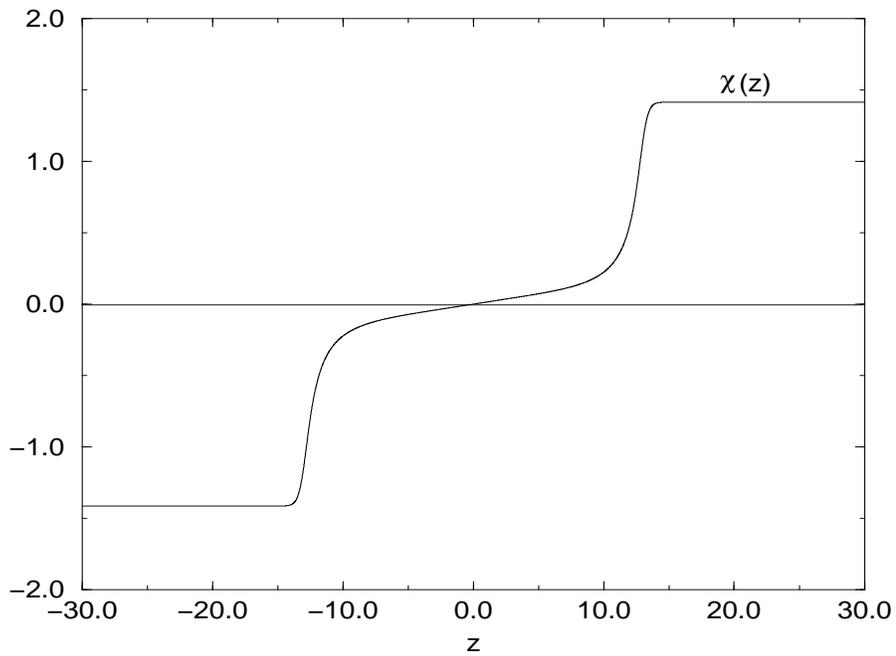,height=4in,width=5.5in}
\caption{\it Profile $\chi(z)$ of a confined-confined domain wall. (a) At high 
temperatures the Coulomb phase is unstable, and one goes directly from one
confined phase to the other. (b) At almost zero temperature the Coulomb phase
is metastable and forms a complete wetting layer that splits the
confined-confined domain wall into a pair of confined-Coulomb interfaces.} 
\end{figure}
Figure 2a corresponds to a high temperature where only the confined phase is 
thermodynamically stable. Then the interface profile interpolates directly
between the two confined phases. Figure 2b corresponds to an almost zero 
temperature where the Coulomb phase is metastable. Then the confined-confined 
domain wall splits into two confined-Coulomb interfaces and the Coulomb phase 
forms a complete wetting layer between them. Figure 3 shows the width $r$ of 
the complete wetting layer as a function of the deviation $\Delta = 
1 - 4ac/b^2$ from phase coexistence. 
\begin{figure}[htb]
\psfig{figure=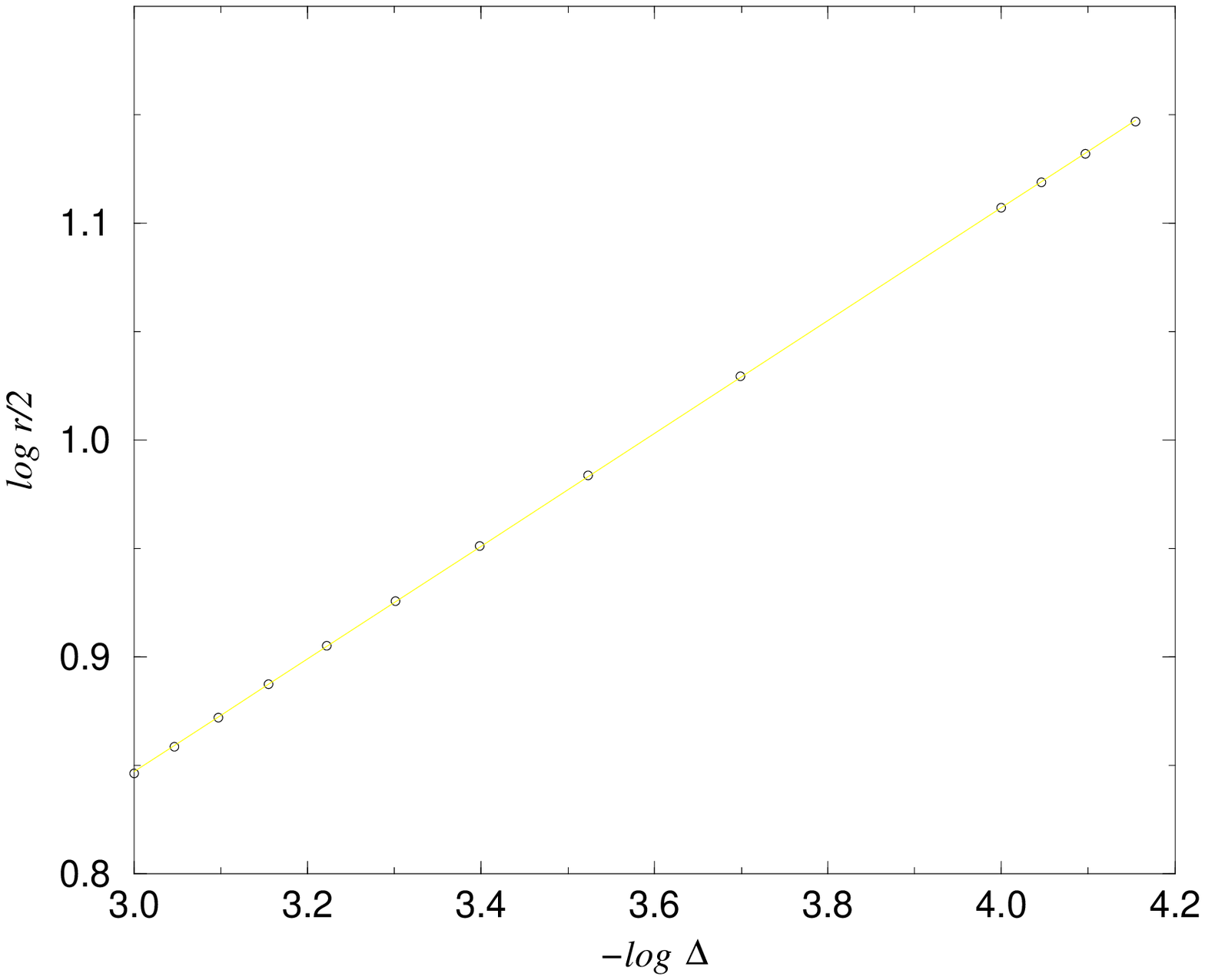,height=4in,width=5.5in}
\caption{\it Determination of the critical exponent $\psi$. The straight line
in the double-logarithmic plot shows the width of the wetting layer as
$r \propto \Delta^{-1/4}$, such that $\psi = 1/4$.}
\end{figure}
One finds $r \propto \Delta^{-1/4}$, such that the critical exponent is 
$\psi = 1/4$. This is the same value that one finds for condensed matter 
interfaces that interact via long-range retarded van der Waals forces with a 
potential $c/r^3$ \cite{Lip84}. We have verified numerically that 
$\alpha_{mn}(T) - 2 \alpha_{0n}(T_c) \propto \Delta^{3/4}$ in agreement with 
the arguments presented in the introduction. The order parameter at the center
of the wetting layer behaves as
\begin{equation}
\frac{d\chi}{dz}(0) \propto \Delta^{1/2},
\end{equation}
which defines another critical exponent.

Let us now turn to $SU(3)$ supersymmetric Yang-Mills theory. In that case, 
complete wetting can occur at zero temperature only if confined-confined
domain walls are not BPS saturated. For the following discussion we will assume
that they are not. The universal aspects of the interface dynamics are again
captured by an effective action
\begin{equation}
S[\chi] = \int d^4 x \ [\frac{1}{2} \partial_\mu \chi^* \partial_\mu \chi +
V(\chi)],
\end{equation}
but the gluino condensate $\chi = \chi_1 + i \chi_2$ is now a complex field. 
The effective potential $V(\chi)$ is restricted by $\Z(3)_\chi$ and charge 
conjugation symmetry. Under chiral transformations $z \in \Z(3)_\chi$, the 
gluino condensate transforms into $\chi' = \chi z$ and under charge conjugation
it gets replaced by its complex conjugate. This implies
\begin{equation}
V(\chi z) = V(\chi), \ V(\chi^*) = V(\chi).
\end{equation}
Again, one needs the potential to be quadratic around the confined phase minima
and quartic around the Coulomb phase minimum. A simple potential with these 
properties is
\begin{equation}
V(\chi) = |\chi|^2(a |\chi|^2 + b \chi_1(\chi_1^2 - 3 \chi_2^2) + c |\chi|^4).
\end{equation}
At zero temperature (corresponding to $b^2 = 4ac$), the above potential has 
four degenerate minima at $\chi=\chi^{(n)}$, $n \in \{1,2,3\}$, 
representing the three confined phases and at $\chi=\chi^{(0)} = 0$ 
representing the Coulomb phase. 

Again, we look for solutions of the classical equations of motion representing
planar domain walls. 
\begin{figure}
\psfig{figure=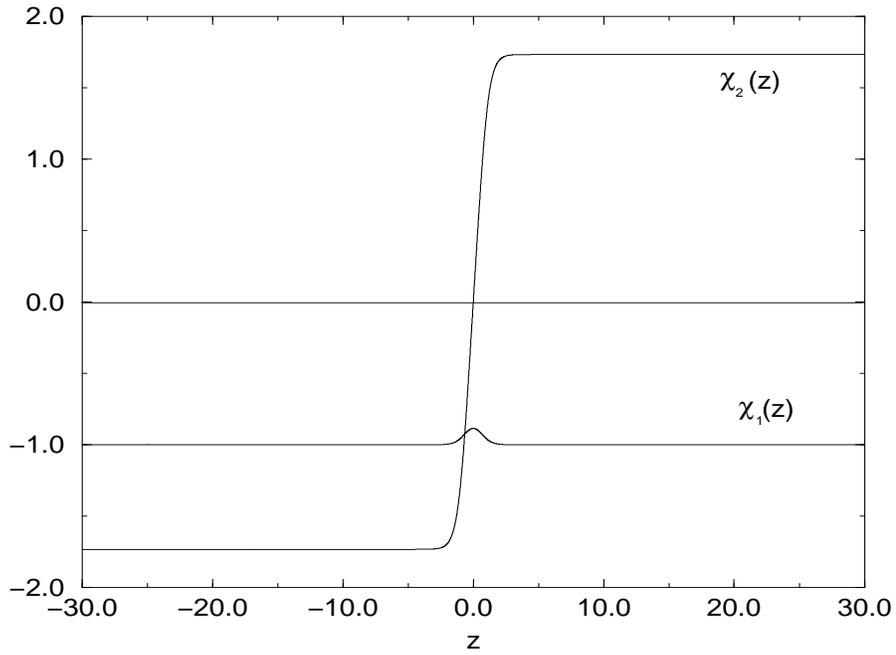,height=4in,width=5.5in}
\psfig{figure=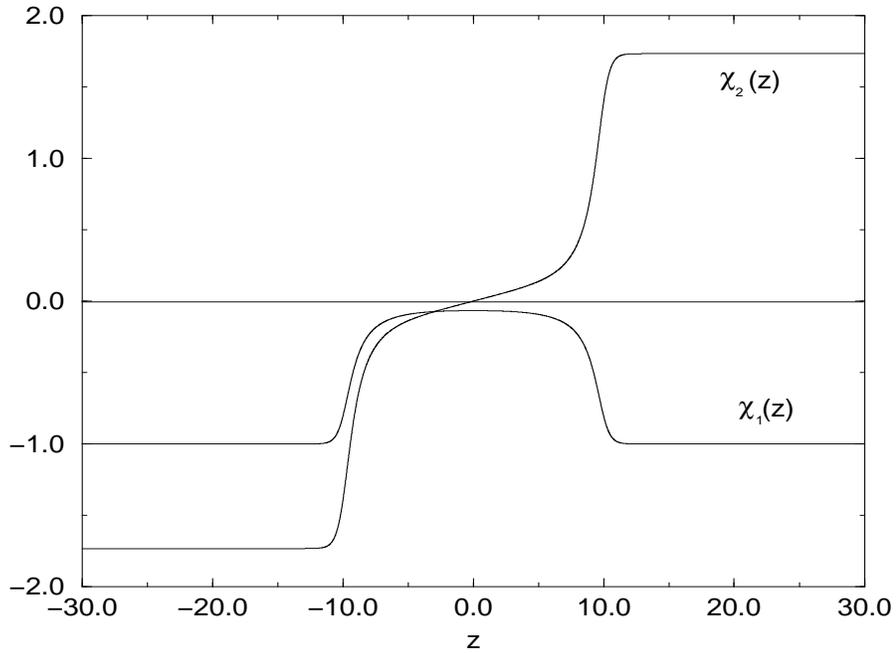,height=4in,width=5.5in}
\caption{\it Shape of a confined-confined domain wall. (a) Deep in the confined
phase the domain wall profile interpolates directly from one confined phase to
the other. (b) At almost zero temperature the wall splits into two 
confined-Coulomb interfaces with a complete wetting layer of Coulomb phase 
between them.}
\end{figure}
Figure 4 shows a numerical solution of these equations for a domain wall 
separating two confined phases with boundary conditions $\chi(\infty) = 
\chi^{(1)}$, $\chi(-\infty) = \chi^{(2)}$. Figure 4a corresponds to a
temperature deep in the confined phase. In this case, the Coulomb phase is
unstable and the domain wall interpolates directly from one confined phase to 
the other. Figure 4b shows the situation at almost zero temperature. Then the 
confined-confined domain wall splits into two confined-Coulomb interfaces and 
the Coulomb phase forms a complete wetting layer between them. 

As in the $SU(2)$ case, we have determined the width of the complete wetting
layer numerically. Again, one finds $r \propto \Delta^{-1/4}$, such that
$\psi = 1/4$. In addition, $\alpha_{mn}(T) - 2 \alpha_{0n}(T_c) \propto 
\Delta^{3/4}$, as it should be. The order parameter at the center of the
wetting layer now behaves as
\begin{equation}
\chi_1(0) \propto \Delta^{1/4}, \ \frac{d\chi_2}{dz}(0) \propto \Delta^{1/2},
\end{equation}
again in agreement with the $SU(2)$ critical exponent. We conclude that the
critical exponents are independent of $N$ and depend only on the range of the
interface interactions. To summarize, at zero temperature complete wetting
occurs for even $N$ and $m-n = N/2$. For odd $N$ or $m-n \neq N/2$ complete
wetting can occur only if confined-confined domain walls are not BPS saturated.
When a confined-confined domain wall is completely wet by the Coulomb phase, 
the complete wetting layer grows with the fourth root of the deviation from the
critical temperature. This is due to long-range interactions in the Coulomb 
phase mediated by massless gluons and gluinos (rather than massive glueballs), 
which lead to $\psi = 1/4$.

\section{Complete Wetting at High Temperature}

Let us now discuss complete wetting at high temperatures where a phase 
transition separates the confined phase from a high-temperature deconfined
phase. We assume that this phase transition is first order, such that confined
and deconfined phases coexist at $T_c$. In the case of ${\cal N} = 0$ 
non-supersymmetric $SU(3)$ Yang-Mills theory at high temperatures the universal
aspects of the interface dynamics are captured by a 3-d effective action 
\cite{Tra92},
\begin{equation}
S[\Phi] = \int d^3 x \ [\frac{1}{2} \partial_i \Phi^* \partial_i \Phi +
V(\Phi)],
\end{equation}
for the Polyakov loop $\Phi$, which is a gauge invariant complex scalar field. 
Its expectation value $\langle \Phi \rangle \propto \exp(- F/T)$, measures the 
free energy $F$ of a static quark. In the confined phase, $F$ diverges and 
$\langle \Phi \rangle$ vanishes, while in the deconfined phase, $F$ is finite 
and $\langle \Phi \rangle$ is non-zero. Under topologically non-trivial gauge 
transformations, which are periodic in Euclidean time up to a center element 
$z \in \Z(3)_c$, the Polyakov loop changes into $\Phi' = \Phi z$. Hence, the 
$\Z(3)_c$ symmetry is spontaneously broken in the deconfined phase. Under 
charge conjugation, the Polyakov loop is replaced by its complex conjugate. The
effective potential $V(\Phi)$ is restricted by $\Z(3)_c$ and charge conjugation
symmetry, i.e.
\begin{equation}
V(\Phi z) = V(\Phi), \ V(\Phi^*) = V(\Phi).
\end{equation}
In this case, all phases are massive (note that the deconfined phase has a 
finite screening length), such that the potential is quadratic around all 
minima. A simple potential with these properties takes the form
\begin{equation}
V(\Phi) = a |\Phi|^2 + b \Phi_1(\Phi_1^2 - 3 \Phi_2^2) + c |\Phi|^4,
\end{equation}
where $\Phi = \Phi_1 + i \Phi_2$. One can restrict oneself to quartic 
potentials because they are sufficient to explore the universal features of the
interface dynamics. At the deconfinement phase transition temperature 
(corresponding to $b^2 = 4ac$), the above potential has four degenerate minima
at $\Phi = \Phi^{(n)}$, $n \in \{1,2,3\}$ representing the three deconfined 
phases and at $\Phi = 0$ representing the confined phase. 
In ref.\cite{Tra92} it 
was shown that a deconfined-deconfined domain wall is completely wet by the 
confined phase and the corresponding critical exponents have been determined 
analytically. As expected for short-range forces, one finds $\psi = 0$, i.e. 
the width of the complete wetting layer diverges logarithmically. In addition, 
for the order parameter at the center of the wetting layer one obtains 
$\Phi_1(0) \propto \Delta^{1/2}$ and $d\Phi_2/dz(0) \propto \Delta^{1/2}$.

In ${\cal N} = 1$ supersymmetric $SU(3)$ Yang-Mills theory the $\Z(3)_\chi$ 
chiral symmetry is spontaneously broken in the confined phase. At high 
temperatures, one expects chiral symmetry to be restored and --- as in the 
non-supersymmetric theory --- the $\Z(3)_c$ center symmetry to be spontaneously
broken due to deconfinement. Consequently, the effective action describing the 
interface dynamics now depends on both order parameters $\Phi$ and $\chi$, such
that
\begin{equation}
S[\Phi,\chi] = \int d^3 x \ [\frac{1}{2} \partial_i \Phi^* \partial_i \Phi + 
\frac{1}{2} \partial_i \chi^* \partial_i \chi + V(\Phi,\chi)].
\end{equation}
The most general quartic potential consistent with $\Z(3)_c$, $\Z(3)_\chi$ and
charge conjugation now takes the form
\begin{equation}
V(\Phi,\chi) = a |\Phi|^2 + b \Phi_1(\Phi_1^2 - 3 \Phi_2^2) + c |\Phi|^4 +
d |\chi|^2 + e \chi_1(\chi_1^2 - 3 \chi_2^2) + f |\chi|^4 + 
g |\Phi|^2 |\chi|^2.
\end{equation}
We assume that deconfinement and chiral symmetry restoration occur at the same 
temperature and that the phase transition is first order. Then three chirally 
broken confined phases coexist with three distinct chirally symmetric 
deconfined phases. The three deconfined phases have $\Phi=\Phi^{(n)}$,
$n \in \{1,2,3\}$, and $\chi = 0$, while the three confined phases are
characterized by $\Phi = 0$ and $\chi=\chi^{(n)}$, $n \in \{1,2,3\}$. 
The phase transition temperature corresponds to a choice of parameters
$a,b,...,g$ such that all six phases represent degenerate absolute
minima of $V(\Phi,\chi)$.

Again, we look for solutions of the classical equations of motion representing
planar domain walls.
\begin{figure}
\psfig{figure=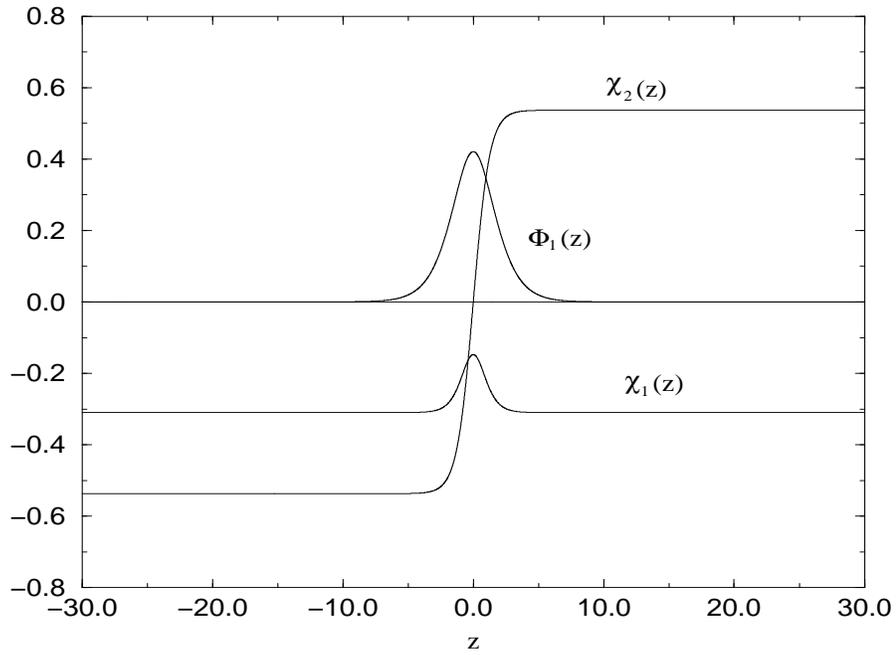,height=4in,width=5.5in}
\psfig{figure=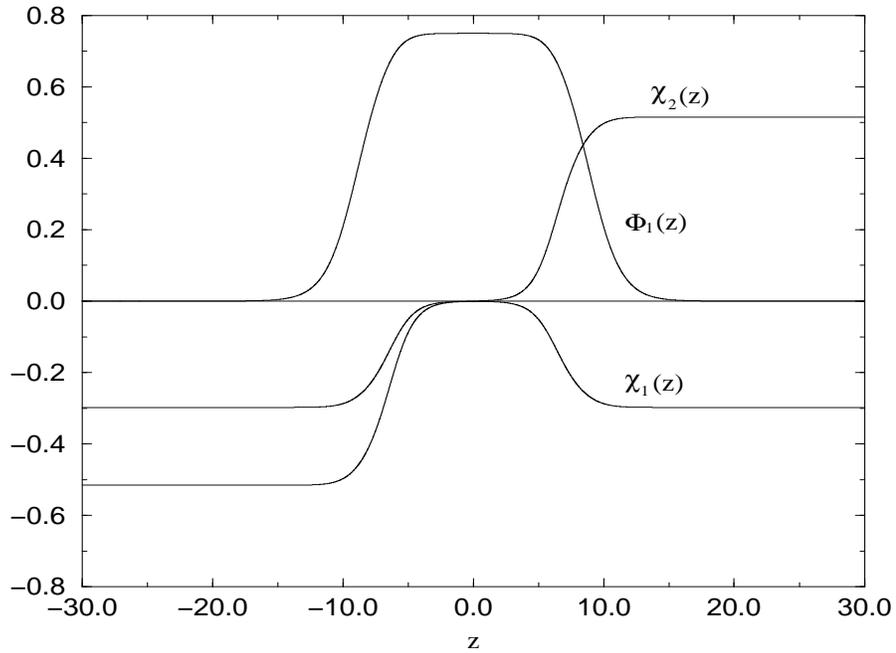,height=4in,width=5.5in}
\caption{\it Shape of a confined-confined domain wall. (a) Deep in the confined
phase $\Phi_1(0) \neq 0$, i.e. the center of the wall has properties of the
deconfined phase. (b) Close to the phase transition the wall splits into two 
confined-deconfined interfaces with a complete wetting layer of deconfined 
phase between them.}
\end{figure}
Figure 5 shows a numerical solution of these equations for a domain wall 
separating two confined phases with boundary conditions $\chi(\infty) = 
\chi^{(1)}$, $\Phi(\infty) = 0$ and $\chi(-\infty) = \chi^{(2)}$, 
$\Phi(-\infty) = 0$. Figure 5a corresponds to a temperature deep in the 
confined phase. Still, at the domain wall the Polyakov loop is non-zero, i.e. 
the center of the domain wall shows characteristic features of the deconfined 
phase. As a consequence, close to the wall the free energy of a static quark is
finite and its string can end there. Thus, as discussed in detail in 
ref.\cite{Cam98}, wetting provides a field theoretic explanation for why QCD
strings can end on domain walls. This effect was first described by Witten in
the framework of M-theory \cite{Wit97}. Figure 5b corresponds to a temperature 
very close to the phase transition. Then the confined-confined domain wall 
splits into two confined-deconfined interfaces and the deconfined phase forms a
complete wetting layer between them. 

For the special values $d = a = 0$, $e = b$, $f = c$, $g = 2 c$ one can find an
analytic solution for a confined-deconfined interface. Combining two of these
solutions to a confined-confined interface, one obtains
\begin{eqnarray}
\Phi_1(z)&=&- \frac{1}{2} \Phi_0 [\tanh\alpha(z-z_0) - \tanh\alpha(z+z_0)], \ 
\Phi_2(z) = 0, \nonumber \\
\chi_1(z)&=&- \frac{1}{4} \chi_0 [2 + \tanh\alpha(z-z_0) - \tanh\alpha(z+z_0)],
\nonumber \\
\chi_2(z)&=&\frac{\sqrt{3}}{4} \chi_0 [\tanh\alpha(z-z_0) + 
\tanh\alpha(z+z_0)],
\end{eqnarray}
where $\Phi_0 = \chi_0 = - 3b/4c$ and $\alpha = - 3b/4\sqrt{c}$. The critical
temperature corresponds to $e^4/f^3 = b^4/c^3$. Near criticality, where
$\Delta = e^4/f^3 - b^4/c^3$ is small, the above solution is valid up to order
$\Delta^{1/2}$, while now $e = b$ and $f = c$ are satisfied to order $\Delta$.
The width of the deconfined complete wetting layer,
\begin{equation}
r = 2 z_0 \propto \log \Delta,
\end{equation}
grows logarithmically as we approach the phase transition temperature. This is 
the expected critical behavior for interfaces with short-range interactions,
which have $\psi = 0$. We have also checked that $\alpha_{mn}(T) - 
2 \alpha_{0n}(T_c) \propto \Delta \log \Delta$, as expected. Again, the order 
parameter at the center of the wetting layer behaves as
\begin{equation}
\chi_1(0) \propto \Delta^{1/2}, \ \frac{d\chi_2}{dz}(0) \propto \Delta^{1/2}.
\end{equation}
Complete wetting also occurs on the other side of the phase transition. Then 
the confined phase completely wets a deconfined-deconfined interface. The
treatment of this case is completely analogous, and the resulting critical
exponents are again those for interfaces with short-range interactions. We
expect the same high-temperature interface critical behavior for other values
of $N$.

\section{Conclusions}

We have investigated universality classes for complete wetting in 
${\cal N} = 1$ supersymmetric $SU(N)$ Yang-Mills theory. At zero temperature,
the Coulomb phase may completely wet a confined-confined interface. In that
case the interfaces interact via long-range forces. The corresponding critical
exponents are the same as those for condensed matter systems with retarded van
der Waals forces with a $c/r^3$ potential. One should keep in mind that we have
assumed that the Coulomb phase indeed exists in supersymmetric Yang-Mills
theory. Should this not to be the case, our calculation, of course, does not 
apply to this theory. Even then, it still correctly describes the effective 
theories discussed in this paper. At high temperatures the massive deconfined 
phase completely wets a confined-confined interface, and the interactions 
between interfaces are short-ranged. Then the width of the complete wetting 
layer grows logarithmically as one approaches $T_c$. Above $T_c$, the confined 
phase can wet a deconfined-deconfined domain wall, with the same critical 
exponents as before. In general, the critical exponents are independent of $N$ 
and depend only on the range of the interface interactions.

\section*{Acknowledgements}

We like to thank R. Lipowsky for an interesting correspondence and A. Smilga
and M. Strassler for very stimulating discussions. The work of A. C. has been 
supported by the Comissionat per a Universitats i Recerca under a cooperative 
agreement between MIT and Generalitat de Catalunya. U.-J. W. wishes to thank 
the A. P. Sloan foundation for its support.


\begin{thebibliography}{10}

\bibitem{Kov97}
A. Kovner, M. Shifman and A. V. Smilga, Phys. Rev. D56 (1997) 7978; \\
G. Dvali and M. Shifman, Phys. Lett. B396 (1997) 64; \\
M. Shifman, M. B. Voloshin, Phys. Rev. D57 (1998) 2590; \\
A. V. Smilga and A. I. Veselov, Nucl. Phys. B515 (1998) 163;\\
G. Dvali and Z. Kakushadze, hep-th/9807140.

\bibitem{Kog98}
I. I. Kogan, A. Kovner and M. Shifman, Phys. Rev. D57 (1998) 5195.

\bibitem{Buf60}
F. R. Buff, in Handbuch der Physik (Springer, Berlin, 1960) Vol. 10,
p. 288.

\bibitem{Wid75}
B. Widom, J. Chem. Phys. 62 (1975) 1332.

\bibitem{Fre89}
Z. Frei and A. Patk\'{o}s, Phys. Lett. B229 (1989) 102.

\bibitem{Tra92}
T. Trappenberg and U.-J. Wiese, Nucl. Phys. B372 (1992) 703.

\bibitem{Cam98}
A. Campos, K. Holland and U.-J. Wiese, hep-th/9805086.

\bibitem{Wit97}
E. Witten, Nucl. Phys. B507 (1997) 658.

\bibitem{Lip84}
R. Lipowsky, Phys. Rev. Lett. 52 (1984) 1429.

\end{thebibliography}
\end{document}